# A Robust and Efficient Trust Management Scheme for Peer-to-Peer Networks

Jaydip Sen

Innovation Lab, Tata Consultancy Services Ltd.,
Bengal Intelligent Park, Salt Lake Electronic Complex, Kolkata – 700091, India
`Jaydip.Sen@tcs.com`

**Abstract.** Studies on the large scale peer-to-peer (P2P) network like Gnutella have shown the presence of large number of free riders. Moreover, the open and decentralized nature of P2P network is exploited by malicious users who distribute unauthentic or harmful contents. Despite the existence of a number of trust management schemes in the literature for combating against free riding and distribution of malicious files, these mechanisms are not scalable due to their high computational, communication and storage overhead. This paper presents a trust management scheme for P2P networks that minimizes distribution of spurious files by a novel topology adaptation scheme. It also reduces search time since most of the queries are resolved within the community of trustworthy peers. Simulation results indicate that the trust management overhead due to the proposed mechanism decreases considerably as the network topology stabilizes.

## 1  Introduction

The term *peer-to-peer* (P2P) system encompasses a broad set of distributed applications which allow sharing of computer resources (e.g. computing power, files, bandwidth etc) by direct exchange between systems. The goal of a P2P system is to aggregate resources available at the edge of the Internet and to share it cooperatively among users. Specially, the file sharing P2P systems have become popular as a new paradigm for information exchange among large number of users in internet scale due to their open, unrestricted, decentralized and anonymous nature.

Depending on the presence of central server, P2P system can be classified as centralized or decentralized [1]. In decentralized architecture, both resource discovery and download are distributed. Decentralized P2P application may be further classified as structured or unstructured network. In structured network, there is restriction regarding the placement of content and network topology. This bounds the number of hops to locate resource. On the other hand in unstructured P2P network, e.g., Gnutella, content of placement is unrelated to topology. Unstructured P2P networks perform better than their structured counterparts in dynamic environment. However, they several problems which need attention such as fake content distribution, free riding (peers who do not share, but consume resources), whitewashing (peers who leave and rejoin the system in order to avoid penalties) etc.

In this paper, a trust-aware, topology adaptation-based efficient searching algorithm is presented that is robust, scalable and light-weight and solves the existing problems in unstructured P2P networks. The rest of the paper is organized as follows. Section 2 discusses some existing related work in the literature. Section 3 presents the proposed algorithm. Section 5 presents the results of simulations carried out on the algorithm. Finally, Section 5 concludes the paper while highlighting some future scope of work.

## 2 Related Work

Several propositions exist in the literature for trust-based searching in unstructured P2P networks. The proposed mechanism in this paper has been motivated from the algorithms proposed in [2] and [3]. In [2], a protocol for the formation of adaptive topologies has been proposed to reduce inauthentic file download and free riding, where a peer connects to those peers from whom it is most likely to download satisfactory content. This strategy is relaxed in [3], where a peer connects to those peers having higher reciprocal capacity. Reciprocal capacity is defined based on peers's capacity of providing good files and of recommending source of download. In [4], a trust-aware adaptive P2P topology is proposed using super peer to control free-riders and malicious peers. Various trust-based incentive schemes have been discussed in [5] to encourage sharing of large number of authentic files. The algorithm proposed in this paper is more efficient and robust than these algorithm since is uses a novel topology-adaptation and trust management scheme as discussed in the subsequent sections.

## 3 The Proposed Trust-Aware Algorithm

This section describes the details of the proposed algorithm. The first subsection presents various design issues involved in the algorithm and the second subsection discusses the operational steps of the algorithm.

### 3.1 Design issues of the proposed algorithm

The various design issues of the algorithm are as follows:

(1)*Network topology and load*: The network is modeled as a *power law graph*, in which the degree distribution of nodes follows power law distribution. The peers adjust topology locally to connect themselves with those peers that provide higher probabilities of supplying authentic files. The network links are categorized into two types: *connectivity link* and *community link*. The connectivity links are the edges of the original power law network. The community links are added between the peers who trust each other. To control bandwidth usage and query processing overhead, an upper limit is put on the additional number of edges that a peer can acquire. If *final_degree(x)* and *initial_degree(x)* be the initial and final degree of peer *x*, the *relative increase in connectivity* (RIC) is bounded by a parameter known as *edge_limit*.

$$RIC = \frac{final\_degree(x)}{initial\_degree(x)} \leq edge\_limit \quad (1)$$

(2) *Content distribution*: It has been shown that Gnutella content distribution follows *zipf distribution* [6]. Therefore, in the proposed algorithm, both content categories and file popularity within each category is modeled with *zipf distribution* with α = 0.8. In this model, each file $f_{c,r}$ is represented by the tuple (*c, r*), where *c* is the content category of the file, and *r* represents its popularity rank within the category. Each content category is characterized by its popularity rank.

(3) *Query initiation model*: The number of queries a peer issues in a simulation round is modeled as a *Poisson* distribution. If *M* is the total number of queries issued in each cycle of simulation and *N* is the number of peers, query rate $\lambda = M/N$ is the mean of the *Poisson* process. The probability that a peer issues query for the file $f_{c,r}$ depends on the peer's interest level in category *c* and the rank *r* of the file.

(4) *Trust management module*: To enable a peer to compute the trustworthiness of other peers, a trust management module is designed. Each peer maintains a *least recently used* (LRU) data structure to keep track of the recent transactions with 32 peers at a time. Every time the peer *i* downloads a file from peer *j*, it rates the transaction as either positive ($tr_{ij}$=1) or negative ($tr_{ij}$=-1) depending on whether downloaded file is authentic or fake. The fraction of successful downloads ($S_{ij}$) that peer *i* had made from peer *j* is recorded by i. If $S_{ij} > 0.5$, *i* considers *j* as trustworthy; if, $S_{ij} < 0$, *j* is considered malicious. If $0 \leq S_{ij} < 0.5$, *i* seeks recommendations from others about *j*.

### 3.2 The proposed trust-aware algorithm

The two major steps of the algorithm are: (i) search and (ii) topology adaptation. These two modules are described in the rest of this section.

(i) *Search*: The searching mechanism used is based on *time to live* (TTL). At each hop, a query is forwarded to a subset of neighbors. The number of neighbors to which the query is forwarded is based an estimate of network connectivity. This estimate is called the probability of community formation, $Prob_{com}$, which for node *x* is computed in (2)

$$Prob_{com} = \frac{degree(x) - initial\_degree(x)}{initial\_degree(x).(edge\_limit - 1)} \quad (2)$$

As simulation proceeds, connectivity of good peers increases and reaches an upper bound. These peers then focus on directing queries to appropriate community where a specific content may exist rather than expanding their communities. The search strat-egy changes from *TTL limited BFS* to *directed DFS*. The search is carried out in two steps– *query initiation* and *query forward* as described below.

*Query initiation*: The source peer forms a query packet containing the name of the file (*c, r*) and forwards it to a subset of neighbors along with $Prob_{com}$ and TTL value. The neighbors are ranked based on both trustworthiness and the similarity of interest. Preference is given to the trusted neighbors sharing similar content categories. Among the trusted neighbors, community members having content matched to the query are preferred. When there is insufficient number of community links, query is forwarded through connectivity links also.

*Query forward*: When a query reaches from peer *i* to peer *j*, following actions are performed: (i) *Check trust level of peer j*: peer *i* checks trust rating of peer *j* through *check trust rating* algorithm (explained later). Accordingly decision of further propagation of the query is taken. (ii) *Check the availability of file*: if the requested file is found, response is sent to peer *j*. If TTL has not expired, the following steps are executed. (iii) *Calculate number of messages to be sent*: the number of messages is computed based on the value of $Prob_{com}$. (iv) *Choose neighbors*: Neighbors are chosen in using *neighbor selection rule*. The search process is shown in Figure 2, in which each query is sent to two neighbors. The matching community links are preferred over connectivity links. Peer *i* initiates query and forwards it to two community neighbors *3* and *4*. The query reaches peer *8* via peer *4*. However, peer *8* knows from its previous experience that peer *4* is malicious. Hence it blocks the query. The query forwarded by peer *5* is also blocked by peer *10* and *11* as both of them know that peer *5* is malicious. Subsequently, the query is matched at four peers: *4*, *6*, *9* and *13*.

(ii) *Topology Adaptation*: Responses are sorted in order of preference by the initiating peer *i* based on the reputation of resource providers. The peer having highest reputation is selected as source of download. The requesting peer checks the authenticity of downloaded file. If the file is found to be fake, peer *i* attempts to download from other sources until it finds the authentic resource. It updates the trust rating and possibly adapts topology to bring trusted peers in its neighborhood and to drop malicious peers from its community. The restructuring of network is controlled by a parameter known as *degree of rewiring*. It represents the probability with which a link is formed between two peers and plays a crucial role in topology adaptation. The adaptation consists of the following operations: (i) *Link deletion*: peer *i* deletes the community link with peer *j* if it finds the later as malicious. (ii) *Link addition*: peer *i* forms community link with peer *j* if the resource is found to be authentic. If $RIC \leq edge\_limit$ for both peers *i* and *j*, then an edge is added subject to the approval of resource provider *j*. If *j* finds peer *i* is malicious, it doesn't approve the link.

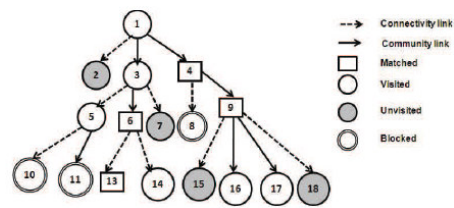

Figure 2. The breadth first search (BFS) tree for the search procedure initiated by peer 1.

## 4  Performance Evaluation

To analyze the performance of the proposed algorithm, following metrics have been used. These metrics are: (i) *attempt ratio* (AR), (ii) *effective attempt ratio* (EAR), and (iii) *query miss ratio* (QMR). AR is the probability that the user downloads the authentic file in the first attempt. The search quality achieved by good peers should be compared to that provided by malicious peers. EAR makes this comparison. If *P(i)* is the total number of attempts made by peer *i* for an authentic file, EAR is given by (3):

$$EAR = (\frac{1}{M}\sum_{i=1}^{M}\frac{1}{P(i)} - \frac{1}{N}\sum_{j=1}^{N}\frac{1}{P(j)})x100 \qquad (3)$$

*M* and *N* are the number of good and malicious peers issuing queries in a particular generation. QMR is defined as the ratio of the number of search failures to the total number of searches in a generation. Due to absence of semantic community initially, number of query misses will be high initially, which will fall later for good peers.

A discrete time simulator written in C is used for simulation. In simulation, 6000 peer nodes, 18000 connectivity edges, 32 content categories are chosen. The degree of deception and the degree of rewiring are taken as 0.1 and 0.3 respectively. The value of the edge_limit is taken as 0.3. The TTL values for BFS and DFS are taken as 5s and 10 s respectively. The discrete time simulator simulates the algorithm repeatedly on the power law network and outputs all the metrics averaged over generations. Barabasi-Alabert generator is used to generate initial power law graphs with 6000 nodes and approximately 18000 edges.

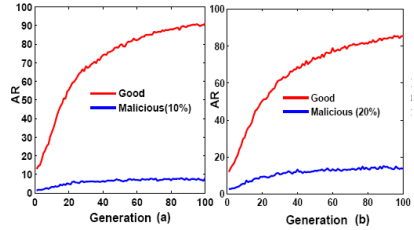

Figure 4. AR vs. percentage of malicious nodes: 10% in (a) and 20% in (b)

Figure 4 shows the cost incurred by each type of peers to download authentic files. It is evident that the cost for downloading an authentic file decreases for a malicious peer as the percentage of malicious peers is increased. It is observed from Figure 5 (a) that with 10% malicious peers, EAR is 80. For good peers, since the queries are forwarded via trusted peers at each hop, there is a high probability of downloading an authentic file in the first attempt. However, the queries from the malicious peers are blocked. As evident from Figure 5, the good peers have higher probability (compared to a malicious peer) of downloading an authentic file in the first attempt as long as the percentage of malicious peers does not exceed 60.

Figure 6 shows QMR for various percentages of malicious peers. Initially, QMR is high as there are no interest-based communities. As the simulation proceeds, the queries are forwarded through the community edges, and the QMR drops for good peers.

Figure 6 shows that steady state value of QMR for good peers is less than 0.2. For malicious peers, QMR is higher since the queries from them are blocked.

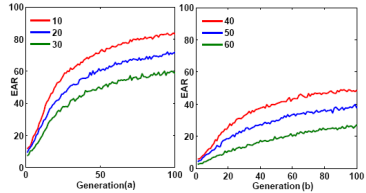

Figure 5. EAR vs. search generation for various percentages of malicious nodes

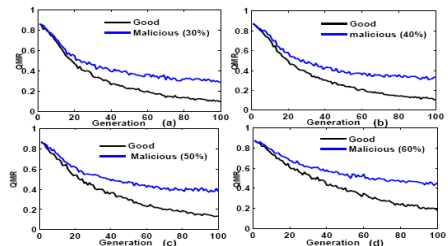

Figure 6. QMR variations with various percentages of malicious peers in the network

## 5  Conclusion

In this paper, an efficient searching mechanism is proposed that solves a number of problems in P2P networks e.g., inauthentic download, free riding and poor search scalability. It has been shown that by judiciously evolving topology, it is possible to isolate the malicious peers while providing high query hit for good peers. The simulation results have shown the efficiency and robustness of the proposed protocol.